\documentclass[useAMS,usenatbib]{mn2e}

\usepackage{natbib}
\usepackage{graphicx}
\usepackage{dcolumn}
\usepackage{tikz}
\usepackage{pgf}
\usepackage{bm}
\usepackage{amsmath}
\usepackage{amssymb}
\usepackage{hyperref}
\usepackage{float}
\usepackage{booktabs}
\usepackage{tabularx}

\graphicspath{{figs/}}

\newcommand{\amina}[1]{{\bf #1}}

\title[The dark halos of the MW's dSphs]{Not too big, not too small:
  the dark halos of the dwarf spheroidals in the Milky Way}

\author[Vera-Ciro et al.]{
\parbox[t]{\textwidth}{
Carlos A. Vera-Ciro\textsuperscript{1}\thanks{E-mail: cavera@astro.rug.nl}, 
Amina Helmi\textsuperscript{1},  Else Starkenburg\textsuperscript{1,2} 
and Maarten A. Breddels\textsuperscript{1}
}\\ 
\\
\\
\textsuperscript{1} Kapteyn Astronomical Institute, Univ. of
Groningen, P.O. Box 800, 9700 AV Groningen, The Netherlands \\
\textsuperscript{2} Dept. of Physics and Astronomy, University of
Victoria, PO Box 3055, STN CSC, Victoria BC V8W 3P6, Canada
}

\begin{document}

\date{\today}

\pagerange{\pageref{firstpage}--\pageref{lastpage}} \pubyear{2012}

\maketitle
\label{firstpage}

\begin{abstract}
  We present a new analysis of the Aquarius simulations done in
  combination with a semi-analytic galaxy formation model. Our goal is
  to establish whether the subhalos present in $\Lambda$CDM
  simulations of Milky Way-like systems could host the dwarf
  spheroidal (dSph) satellites of our Galaxy. Our analysis shows that,
  contrary to what has been assumed in most previous work, the mass
  profiles of subhalos are generally not well fit by NFW models but
  that Einasto profiles are preferred. We find that for shape
  parameters $\alpha = 0.2 - 0.5$ and $v_{\rm max} = 10 - 30$ km/s
  there is very good correspondence with the observational constraints
  obtained for the nine brightest dSph of the Milky Way. However, to
  explain the internal dynamics of these systems as well as the number
  of objects of a given circular velocity the total mass of the Milky
  Way should be $\sim 8 \times 10^{11}$~M$_{\odot}$, a value that is
  in agreement with many recent determinations, and at the low mass
  end of the range explored by the Aquarius simulations. Our
  simulations show important scatter in the number of bright
  satellites, even when the Aquarius Milky Way-like hosts are scaled
  to a common mass, and we find no evidence for a missing population
  of massive subhalos in the Galaxy. This conclusion is also supported
  when we examine the dynamics of the satellites of M31.
\end{abstract}


\section{Introduction}
\label{sec:intro}

Despite the great success of the $\Lambda$CDM concordance cosmological
model on large scales, on the scales of galaxies and below the theory
is often defied. Some of the issues on small scales have been
consistently explained within the theory itself with the inclusion of
physical processes that mostly affect baryons. This is the case for
the ``missing satellite problem'' \citep{Klypin1999, Moore1999},
namely the overabundance of satellites in dark matter only simulations
compared to the observed number of luminous objects around the Milky
Way and other nearby galaxies. It is now widely accepted that the
shallow potential wells of small dark matter halos must be strongly
affected by reionization and feedback, making star formation highly
inefficient in such systems \citep{Couchman1986, Efstathiou1992,
  Kauffmann1993, Thoul1996, Bullock2000, Somerville2002, Benson2002b,
  Li2009, Okamoto2009, Maccio2010, Stringer2010, Font2011, Guo2011}.

Recently, \citet{Boylan2011} have argued that the dark matter
satellites (subhalos hereafter) predicted by $\Lambda$CDM are
persistently too dense to host the observed population of dwarf
spheroidal galaxies (dSph) in the Milky Way if these are embedded in
halos following \citet[][hereafter NFW]{nfw1996, nfw1997}
profiles. More recently, \citet{Boylan2011b} presented an even
stronger argument (free of the assumption of a specific density
profile) and argued that the Milky Way is missing a population of
massive satellites. A few studies have been published in the
literature that address this conundrum. \citet{Lovell2011} showed that
in warm dark matter cosmological simulations of Milky Way-like halos,
the circular velocity curves of subhalos are consistent with the
constraints derived by \citet{Wolf2010} for the Milky Way satellites
\citep[see also][]{Walker2009}. Following a similar line,
\citet{Vogelsberger2012} carried out simulations of self-interacting
dark matter and showed that the most massive subhalos develop cores,
what could partially solve the problem. On the other hand,
\citet{DiCintio2011} pointed out that by including baryons in the cold
dark-matter context the problem becomes more severe probably due to
the additional adiabatic contraction experienced by the dark matter
subhalos hosting gas.

There are two assumptions sometimes implicit in the models which may
lead to biased answers if overlooked. These concern (i) the actual
mass of the Milky Way\footnote{In fact, shortly after we submitted our
  manuscript for publication, \cite{Wang2012} analysed the Millennium
  Simulation series and used the invariance of the scaled subhalo
  velocity function to argue that the absence of massive subhaloes
  might indicate that the MW is less massive than commonly assumed.}
and, (ii) the density profiles followed by dark matter satellites
assembled in $\Lambda$CDM. The first issue has been addressed with a
plethora of methods leading to measures that, usually, are consistent
with a total mass of $0.7 - 2.0 \times 10^{12}$ M$_{\odot}$
\citep{Kochanek1996, Wilkinson1999, Sakamoto2003, Battaglia2005,
  Battaglia2006, Smith2007, Li2008, Xue2008, Kallivayalil2009,
  Guo2010, Watkins2010, Gnedin2010, Busha2011}. The measurements
suffer from uncertainties in the modeling as well as limitations in
the kinematics of the tracers used. Therefore comparisons to
simulations of Milky Way dark matter halos should take into account
this uncertainty.

On the second issue, namely the density profile of subhalos,
significant progress has been made, especially in recent years.
Already \citet{Stoehr2002} found that the circular velocity curves of
subhalos in cosmological $N$-body simulations are more narrowly peaked
(in a log-log plot) than the widely used NFW models and explored how
consistent these were with the internal kinematics of the Milky Way
satellites.  The outstanding numerical resolution achieved in the
latest of such cosmological $N$-body simulations has enabled a closer
examination of the shape of the density profile down to the innermost
few parsecs of dark matter halos \citep{Springel2008, Madau2008}. Such
studies have shown that Einasto models provide better matches to the
density profiles found in the simulations than the NFW form
\citep{Navarro2010, Reed2011, DiCintio2012}, confirming previous
results on the subject \citep{Navarro2004, Merrit2005, Merrit2006,
  Graham2006, Prada2006, Gao2008}.

In this paper, we will reconsider both these issues and establish how
much they affect the conclusions drawn by
\cite{Boylan2011,Boylan2011b}. Like these authors we use the Aquarius
halos, but we supplement the dynamical information provided by the
simulations with a semi-analytic galaxy formation model
\citep{Starkenburg2011}. One of the advantages of this approach is
that it enables us to directly compare objects in the simulation with
those observed. We introduce some relevant features of the simulations
in Section \ref{sec:simulations}, while in Section \ref{sec:results}
we present in detail the results of our analysis. We draw our
conclusions in Section \ref{sec:conclusions}.

\section{Numerical Preliminaries}
\label{sec:simulations}

\begin{figure*}
  \includegraphics[width=0.95\textwidth]{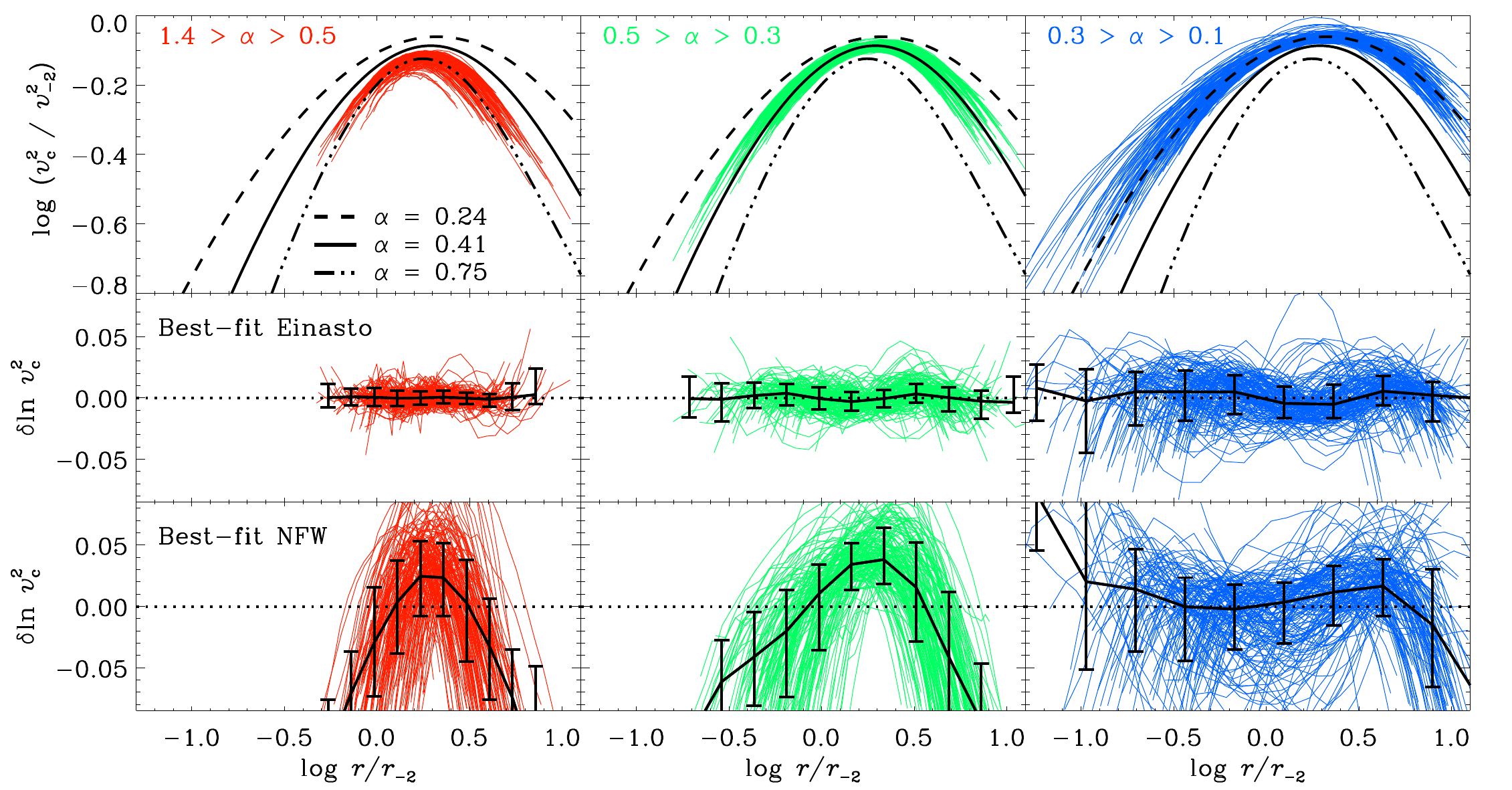}
  \caption{Spherically averaged circular velocity profiles $v_c^2 (r)
    = Gm(r)/r$ for the subhalos that are predicted to host stars by
    our semi-analytic model. Velocities have been scaled to $v_{-2}^2
    \equiv 4\pi G \rho_{-2} r_{-2}^2$. As already reported in
    \citet{Stoehr2002} the velocity profiles of subhalos tend to be
    more narrowly peaked than in the NFW form. The sample of subhalos
    has been grouped according to the best fit value of $\alpha$, and
    plotted with different colors. The number of objects in each
    $\alpha$-bin is 154. An Einasto profile with the average value of
    $\alpha$ for each bin is overplotted, while the median $v_{\rm
      max}$ in each $\alpha$-bin is 13.6, 20.3 and 27 km~s$^{-1}$ from
    left to right, and the $v_{\rm max}$ ranges given by the 68\%
    percentiles for each panel are (10, 26.2), (15.9, 26.2) and (20.5,
    44.8) km~s$^{-1}$, respectively.  The residuals from the
    best-Einasto (NFW) fits are shown in the middle (bottom) panel,
    and in general are consistent with zero for the Einasto profile
    and exhibit systematic deviations from zero for the NFW case. In
    the column $\langle\alpha\rangle = 0.24$ both models yield similar
    results, which is naturally expected since the NFW equivalent is
    reached with $\alpha=0.22$. The systematic change of $r_{\rm
      conv}/r_{-2}$ with $\alpha$ is a consequence of setting the
    convergence parameter $\kappa$ to a fixed value.  }
\label{fig:einasto-fits}
\end{figure*}

We use the simulations of the Aquarius project, six Milky Way-sized
dark matter halos assembled in a background cosmology consistent with
the constraints yielded by WMAP-1. Each halo (labeled from \texttt{A}
to \texttt{F}) was simulated at different resolutions, starting from a
particle mass $m_p = 3.143\times 10^{6}$ M$_{\odot}$ for the lowest
and $m_p = 1.712\times 10^{3}$ M$_{\odot}$ for the highest
resolution. In what follows we focus on the level \texttt{2} which is
the highest level at which {\it all} the Aquarius halos were simulated
\citep[For more details see][]{Springel2008}.

Subhalos in these simulations are identified as bound overdensities
with \textsc{subfind} \citep{Springel2001}. For each subhalo we
compute the circular velocity profile as $v_c^2(r) = Gm(r)/r$, where
$m(r)$ is the mass enclosed within the spherical radius $r$. The
maximum circular velocity $v_{\rm max}$ is defined as the peak of the
circular velocity curve, and is reached at position $r_{\rm max}$,
i.e. $v_c(r_{\rm max}) = v_{\rm max}$.

Numerical convergence is established by looking at the convergence
radius as defined by \citet{Power2003}. \cite{Navarro2010} showed that
the roots of the equation
\begin{equation}
  \kappa = \frac{\sqrt{200}}{8} \frac{n(r)}{\ln
    n(r)}\left[\frac{\overline{\rho}(r)}{\rho_c}\right]^{-1/2},
\end{equation}
correspond to different degrees of convergence depending on the value
of the parameter $\kappa$. In this equation $n(r)$ is the number of
particles enclosed within the radius $r$ and
$\overline{\rho}(r)/\rho_c$ is the spherical density at this position
in units of the critical value. Comparing the various resolutions of
the Aquarius simulations, \citet{Navarro2010} showed that $\kappa =
(7.0,1.0,0.4)$ correspond to deviations in the circular velocity
profile of about $(2.5\%,10\%,15\%)$ respectively. Here, we use
$r_{\rm conv} (\kappa = 0.4)$ for each subhalo. We also define the
tidal radius $r_{\rm tidal}$ of a subhalo as that which encompasses
$95\%$ of the bound particles. The results from our definition show
good agreement with the output from more sophisticated expressions for
the tidal radius \citep{Tormen1998}.

To make more direct comparisons to the satellite population of the
Milky Way, we have also run a semi-analytic model of galaxy formation
for all the Aquarius halos. This model is based on that originally
developed by \cite{Kauffman1999, Springel2001, DeLucia2004,
  Croton2006, DeLucia2007} and later modified to describe more
accurately processes on the scales of dwarf galaxies
\citep{Li2010}. The implementation used here also includes recipes for
stellar stripping and tidal disruption. The resulting satellite
luminosity function agrees well with that of the Milky Way as reported
by \cite{Koposov2008} (see Section \ref{sec:results-SA}). Also the
internal properties of the satellites, such as scaling relations,
metallicities and star formation histories are in good agreement with
those observed \citep[for more details see][]{Starkenburg2011}.

\section{Results}
\label{sec:results}

\subsection{About the density profiles}

It is has been already reported in the literature that the mass
profiles of $\Lambda$CDM halos deviate from the NFW functional form
\citep{Stoehr2002, Navarro2004, Merrit2005, Merrit2006, Graham2006,
  Prada2006, Gao2008}. Using the Aquarius simulations,
\cite{Navarro2010} showed that a parametric model with a density
profile with logarithmic slope described by a power-law (Einasto
profile) provides better fits for objects of virial mass $\sim
10^{12}$ M$_{\odot}$. The power index $\alpha$ adds another free
parameter, therefore the fits are expected to improve. Nevertheless,
it was shown by \cite{Springel2008} that even after fixing
$\alpha=0.16$ the Einasto profile still yields much better
results. The nature of the shape parameter $\alpha$ has been recently
investigated for isolated objects with masses in excess of $5\times
10^{12}$ M$_{\odot}$ and the results suggest a deep connection with
the pseudo-phase-space density distribution \citep{Ludlow2011}. For
subhalos \cite{Springel2008} also showed that the mass profiles follow
much closer the Einasto than the NFW model. For both models, the mass
enclosed within the spherical radius $r$ can be written as
\begin{subequations}\label{eq:mass}
\begin{equation}\label{eq:mass-all}
  m(r) = 4\pi r_{-2}^3\rho_{-2} g(r/r_{-2}),
\end{equation}
where $r_{-2}$ is the radius at which the logarithmic slope of the
density profile reaches the isothermal value and $\rho_{-2}$ is the
density at that position. The details of each model are inherited by
the function $g$, which takes the form
\begin{equation}\label{eq:mass-nfw}
  g_{\rm NFW}(x) = 4\ln (1 + x) - \frac{4x}{1+x},
\end{equation}
for the NFW profile, and
\begin{equation}\label{eq:mass-einasto}
  g_{\rm Einasto}(x) = \frac{1}{\alpha}\exp\left( \frac{3\ln\alpha + 2
  - \ln 8}{\alpha}\right) \gamma \left(\frac{3}{\alpha},
  \frac{2x^\alpha}{\alpha} \right),
\end{equation}
\end{subequations}
for the Einasto model. Here $\gamma (a, x)$ is the lower incomplete
gamma function. Although intrinsically different, these profiles
resemble each other for $\alpha\approx0.22$ in $0.01\leq r/r_{-2} \leq
100$. That means that objects that have a shape parameter close to
this value are well fitted by either
model. Fig. \ref{fig:einasto-fits} shows the spherically averaged
circular velocity profiles $v_c^2 = Gm(r)/r$ for all the subhalos that
are predicted to host stars according to our semi-analytical model. In
total we calculate 20 bins in the region $r_{\rm conv} \leq r \leq
0.9\, r_{\rm tidal}$, we use this upper cutoff to ensure that our fits
are not driven by tidal effects. All objects have at least 200
particles, but generally significantly more than 1000. For each of the
plotted subhalos we calculate the merit function
\begin{equation} \label{eq:chi2}
E = \frac{1}{N_{\rm bins}}\sum_{i = 1}^{N_{\rm bins}} (\ln
v_{c}^2(r_i) - \ln v_{c,i}^2)^2,
\end{equation}
and minimize it against the free parameters of each model.  We have
deliberately chosen to use the cumulative mass instead of the
differential profile since it is less sensitive to the shot-noise of
each bin, as a consequence we can go to low number of particles,
whenever the restriction $n(0.9\,r_{\rm tidal}) - n(r_{\rm conv}) \geq
200$ is met.

Fig. \ref{fig:einasto-fits} shows the results of our fitting
procedure. Here the subhalos have been distributed in bins with
  equal number of objects (namely 154), and according to their best
fit $\alpha$ value. The three different columns show the subhalos that
fall into each $\alpha$-bin, the average $\alpha$ within each bin is
quoted in the top-left panel. Each curve in the top row has been
conveniently normalized to a characteristic velocity $v_{-2}^2 \equiv
4\pi G \rho_{-2} r_{-2}^2$ and the characteristic radius $r_{-2}$.  We
have also overplotted the predicted Einasto profiles for the average
$\alpha$. The middle panels show the residuals of the best-Einasto fit
for each subhalo, the thick line represent the median and $1\sigma$
equivalent dispersion. The residuals are consistent with zero
indicating that the Einasto profile fits better than NFW (whose
residuals are shown in the bottom panel), especially for large
$\alpha$ values. Interestingly, for $\langle\alpha\rangle = 0.24$ the
NFW model provides a good and comparable fit to the Einasto model (see
bottom-right panel). This is actually expected, since $\alpha=0.22$
represents a model that nearly follows the NFW profile.

On average lower mass subhalos tend to have larger values of
$\alpha$. However this correlation has a large scatter, for instance
the maximum circular velocity for objects in the central panel of
Fig. \ref{fig:einasto-fits}, i.e. those with $0.3< \alpha < 0.5$ is in
the range $10.5 < v_{\rm max}/{\rm km\;s^{-1}} < 48.2$.

\subsection{MW's dSphs constraints revisited}

\begin{figure}
  \includegraphics[width=0.48\textwidth]{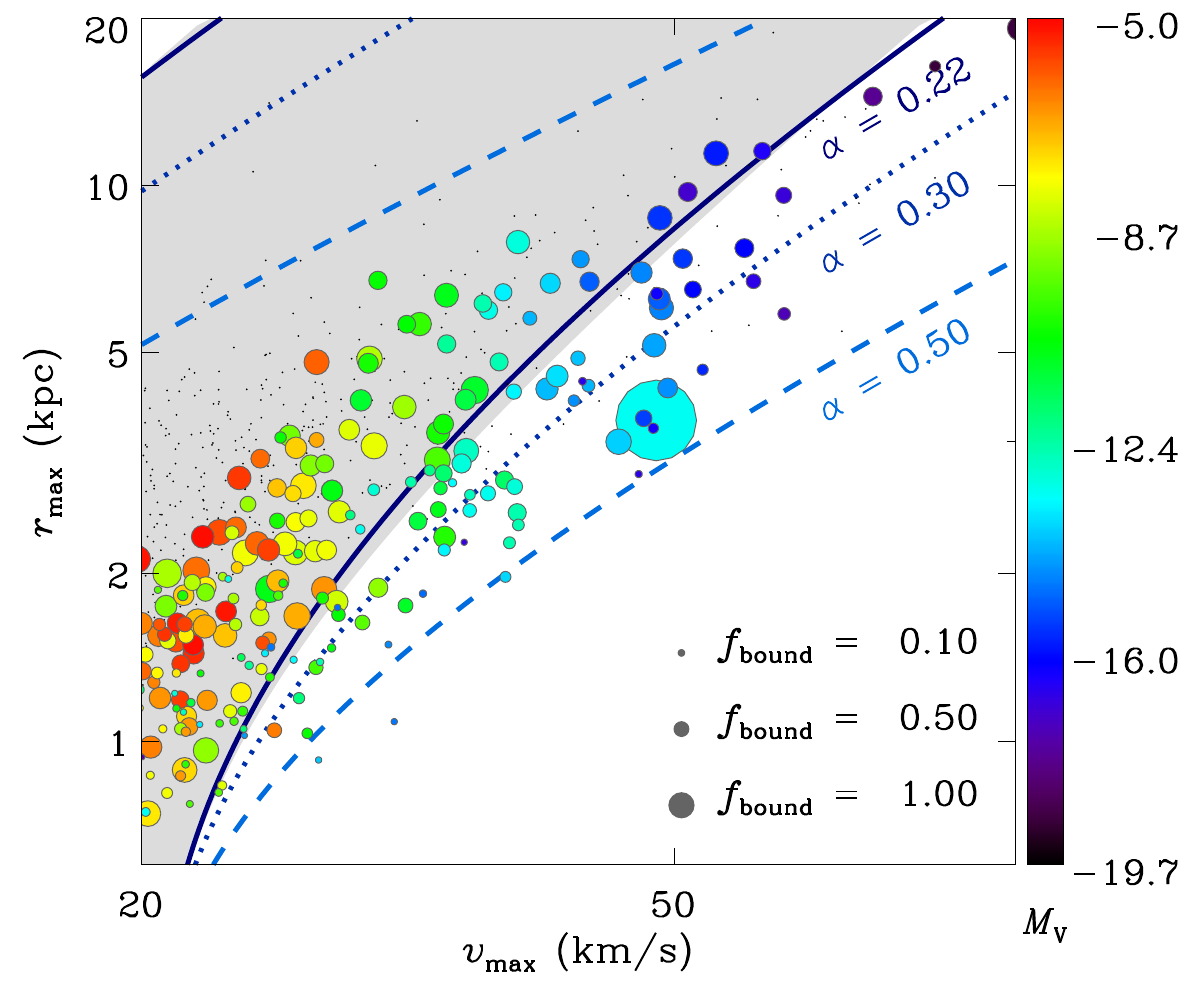}
  \caption{Constraints for the MW's dwarf spheroidals using NFW (gray
    band) and Einasto (blue curves) profiles. Points are the results
    from the six Aquarius halos, colored according to their predicted
    luminosity and sized using the fraction of mass retained after
    infall. The cyan point at $v_{\rm max} \sim 50$ km s$^{-1}$
    represents a subhalo that underwent a merger with another
    substructure after infall, therefore increasing its mass. The
    black dots correspond to isolated halos in the simulations.}
\label{fig:rmax-vmax}
\end{figure}

\citet{Wolf2010} have shown that the mass enclosed within the half
light radius $r_{1/2}$ of a dynamical system can be robustly
determined as $m_{1/2} = 3G^{-1}\langle \sigma_v^2\rangle r_{1/2}$,
without (precise) knowledge of its velocity anisotropy. Here $\langle
\sigma_v^2\rangle$ is the light-weighted average line-of-sight
velocity dispersion of the system.  In the case of dSph galaxies, this
effectively implies a measurement of the circular velocity at
$r_{1/2}$, which therefore constrains the possible family of circular
velocity curves for a given dark matter density profile. Following
\citet{Boylan2011} we plot in Fig. \ref{fig:rmax-vmax} the $2\sigma$
constraints derived in this way for the 9 most luminous dSphs
satellites of the MW (i.e. excluding the Sagittarius dwarf galaxy and
the Small and Large Magellanic Clouds). Like \citet{Boylan2011} here
we assume these systems are embedded in NFW profiles, which leads to
the gray band shown in the figure.  This band is also nearly
consistent with the masses enclosed within 300 pc as reported by
\citep{Strigari2008} $2.5\times 10^6 \leq m(300\;{\rm
  pc})/\rm{M}_{\odot} \leq 3.0\times 10^7$.

The advantage of using this plot is that one can directly compare
against the results extracted from $\Lambda$CDM simulations. The
filled circles in Fig. \ref{fig:rmax-vmax} show the distribution of
$(r_{\rm max}, v_{\rm max})$ measured directly from our simulations
for the satellites hosting stars. The colors indicate the predicted
luminosities and the sizes correspond to the bound mass fraction at
present day, i.e. $f_{\rm bound} = m(0) / m(z_{\rm infall})$, where
$z_{\rm infall}$ is the lowest redshift at which the progenitor of a
given subhalo was not associated to one of the main Aquarius halos.
Here, the values $(r_{\rm max}, v_{\rm max})$ have been corrected for
softening length effects following the expressions given by
\cite{Zavala2010}. As highlighted in the Introduction, there are
important differences in the location of the points from the
simulations and those derived for the dSph satellites of the Milky
Way, which may lead to the conclusion that there is a significant
problem with our currently preferred cosmological model.

However, this comparison may be need to be revisited since we
demonstrated in the previous section that an NFW profile is not
expected to describe well the dark matter halos of satellites in
$\Lambda$CDM. Therefore, we have computed the family of $(r_{\rm max},
v_{\rm max})$ values that are consistent with the measurement of
$v_c(r_{1/2})$ for the dSph of the Milky Way, but now we have
considered Einasto profiles.  The $2\sigma$ constraints are shown in
Fig. \ref{fig:rmax-vmax}. Given the freedom we have in choosing the
extra parameter $\alpha$, we have plotted 3 different bands
corresponding to $\alpha = (0.22,\; 0.30,\; 0.50)$. The $\alpha=0.22$
(solid lines) is consistent with the NFW predictions, as expected. For
values of $\alpha\lesssim 0.5$ (dashed line) the constraints from
observations actually overlap with those found in the
simulations. This range of values of $\alpha$ lies well within
  the range observed in Fig.~\ref{fig:einasto-fits}, since $\sim 2/3$
  of our sample has $\alpha < 0.5$. 

From Fig. \ref{fig:rmax-vmax} we note that there is a correlation
between the value of $\alpha$ and $f_{\rm bound}$ i.e.\ the amount of
stripping a subhalo has experienced. Very heavily stripped objects
require, on average, higher $\alpha$, and deviate the most from NFW
profiles.  The black dots in this figure correspond to the location of
isolated dark matter halos in the same $v_{\rm max}$ range as the
satellites. This confirms that such isolated dark matter halos are
well-fit by NFW profiles, and that tidal stripping is acting on the
subhalos to change the shape of their circular velocity profile to the
Einasto form \citep[see also][]{Hayashi2003}.

The mismatch between the observations \citep[with the assumption of
NFW,][]{Boylan2011} and the simulations is only partly alleviated with
the use of an Einasto profile as Fig.~\ref{fig:rmax-vmax} shows. The
Milky Way does not have many very luminous dwarf galaxy
satellites. Brighter than Fornax ($M_V \sim -13.2$), only the
Sagittarius dwarf and the Magellanic Clouds are known, and none is
included in Fig. \ref{fig:rmax-vmax}. According to our semi-analytic
model such luminous objects would populate the upper-right of this
plot, i.e. $v_{\rm max} \gtrsim 40$ km/s and $r_{\rm max} \gtrsim 2.5$
kpc.  Therefore, in this region of the diagram, the mismatch between
the observations and the simulations {\it has} to be entirely
attributed to the absence of other bright dSph in the Milky Way, which
is the point originally raised by \cite{Boylan2011}. In other words,
the lack of objects brighter (or more massive) than Fornax around the
Milky Way cannot be explained away through a change in the density
profile of the dark matter subhalos in the context of the $\Lambda$CDM
model, unless these have been very heavily stripped. However, such
massive subhalos are typically accreted late, and hence have not
suffered significant amounts of stripping.  It is the region with
$v_{\rm max} \sim 20 - 40$ km/s, where we expect to find the subhalos
hosting most of the classical dSph according to this plot, where the
difference between assuming an NFW or an Einasto profile needs to be
taken into account to bring the simulations in agreement with the
observations.


\subsection{Effects of the host halo mass}
\label{sec:results-SA}

\begin{figure}
  \includegraphics[width=0.48\textwidth]{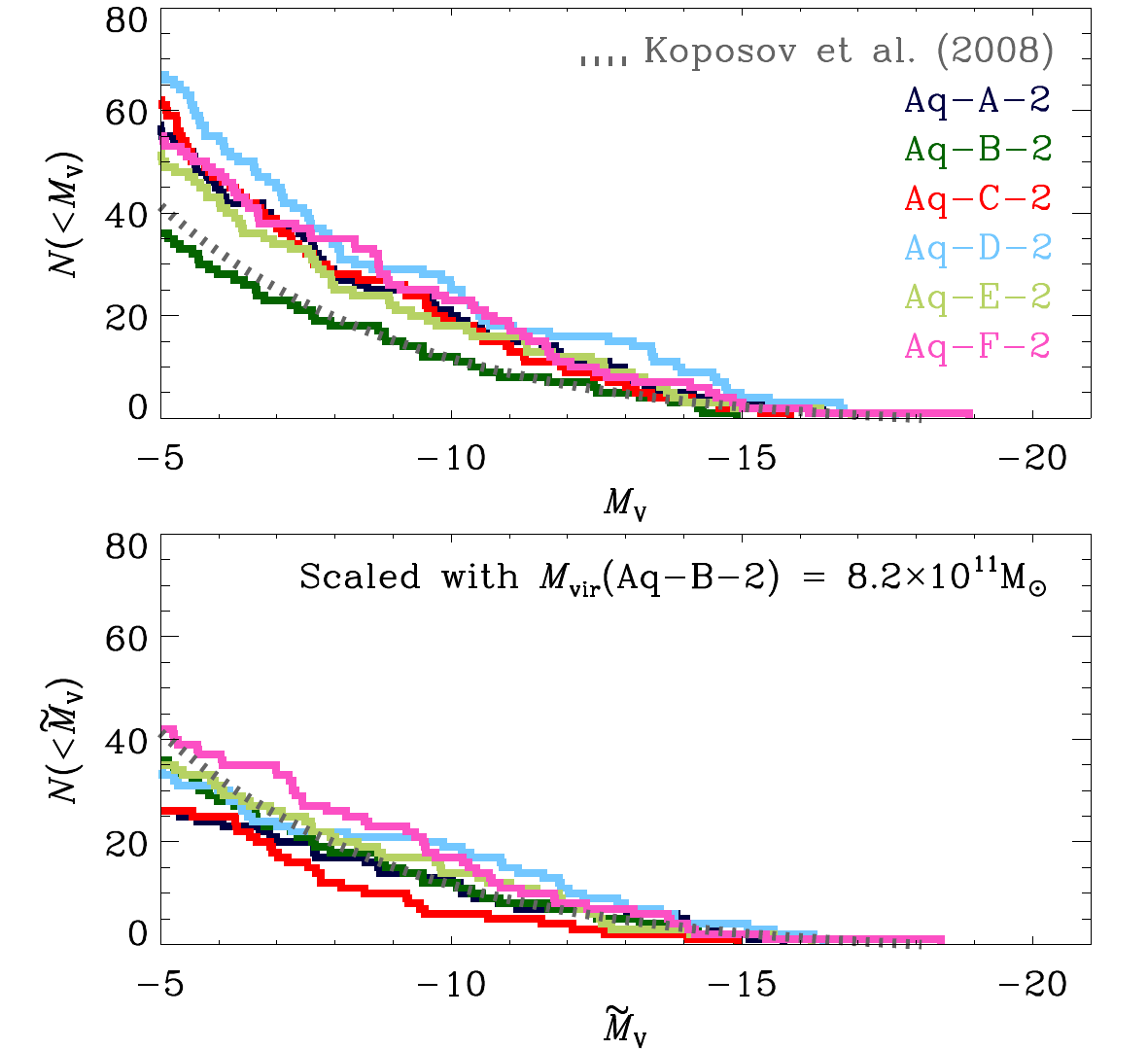}
  \caption{ Luminosity function for the original Aquarius simulations
    (top) and once they have scaled to the mass of \texttt{Aq-B-2}
    (bottom). For reference we have added the luminosity function
    derived by \citet{Koposov2008} for the Milky Way. This takes into
    account incompleteness issues for satellites with $M_V > -11$ and,
    for brighter objects it considers the average for the Milky Way
    and M31. Although in the scaled version the simulations follow
    much more closely the observations, some differences remain in the
    number of satellites of a given luminosity.}
\label{fig:lum-fn}
\end{figure}

\citet{Springel2008} have shown that the mass function of dark matter
halos is independent of mass, i.e. that it is self-similar. This
implies that the number of subhalos of a given mass scales directly
with the mass of the host \citep[although][suggest that the slope is
slightly larger for $10^{15}$ M$_{\odot}$
objects]{Gao2012}. Therefore, we can expect that, down to a certain
scale, brighter or more massive central galaxies will host a larger
number of satellites. This is indeed shown in the top panel
Fig. \ref{fig:lum-fn}, where we have plotted the luminosity function
of all Aquarius halos. It is clear from this figure that this is the
case, since the three heavier of the Aquarius halos \texttt{Aq-A-2},
\texttt{Aq-C-2} and \texttt{Aq-D-2} have 57, 62 and 67 satellites
respectively, while the lightest, \texttt{Aq-B-2} has only 36
satellites with $M_V \leq -5$, and hence has the shallowest luminosity
function in the faint end. It is possible to show that a doubling of
the mass of the host halo roughly leads to an increase by a factor of
$\sim 2$ in the number of satellites brighter than $M_V = -5$
\citep{Starkenburg2011}.

We thus explore the effect of host halo mass on the properties of our
simulated satellites, by re-scaling all halos to a common value
following \cite{Helmi2003}. Because of the scale-free nature of
gravity, we may assume that if a halo of mass $M_{\rm Aq}$ is scaled
to have a mass $M_{\rm MW}$ then the subhalos's masses $m$ should be
scaled as
\begin{subequations}\label{eq:scale}
\begin{equation} \label{eq:scale-m}
\widetilde{m} = m \frac{M_{\rm MW}}{M_{\rm Aq}} \equiv \mu m.
\end{equation}
Naturally the distances will become
\begin{equation}\label{eq:scale-r}
  \widetilde{r} = \mu^{1/3}r,
\end{equation}
while the circular velocity profiles 
\begin{equation}\label{eq:scale-v}
  \widetilde{v_c} = \left(\frac{G
    \widetilde{m}}{\widetilde{r}}\right)^{1/2} = \mu^{1/3}v_c.
\end{equation}
\end{subequations}
To determine the factor $\mu \equiv M_{\rm MW}/ M_{\rm Aq}$, we need
to specify $M_{\rm MW}$. As discussed in the Introduction, the value
of the total mass of the Milky Way is quite uncertain. However,
motivated by the remarkable match between the luminosity function of
\texttt{Aq-B-2} and that of the Milky Way \citep{Koposov2008}, we set
$M_{\rm MW} = M_{\rm vir} (\texttt{Aq-B-2}) = 8.2\times 10^{11} {\rm
  M}_\odot$\footnote{In this paper we denote $M_{\rm vir} = M_{200}$,
  i.e. the the mass enclosed in a sphere with mean density 200 times
  the critical value.}. This value is consistent with many recent
studies using different techniques \citep[e.g.][]{Battaglia2005,
  Battaglia2006, Smith2007, Xue2008}. This implies that the value of
$\mu$ ranges from unity to $2.2$ at most, which implies that distances
and velocities in the scaled simulations will be at most decreased by
a factor of $1.3$.

We run our semi-analytic galaxy formation model now for the re-scaled
Aquarius simulations.  The resulting luminosity function is shown in
the bottom panel of Fig. \ref{fig:lum-fn}, where the new predicted
magnitudes are denoted by $\widetilde{M}_V$. It is evident from this
figure that each halo now follows much more closely the Milky Way's
luminosity function. It is important to note that there is still some
halo-to-halo dispersion, which can be attributed to the stochastic
nature of the mass assembly of each object. That is, not all the
galaxies with the same mass are expected to have the same number of
satellites with the same luminosity, although some form of statistical
equivalence should be present.

\begin{figure}
  \includegraphics[width=0.48\textwidth]{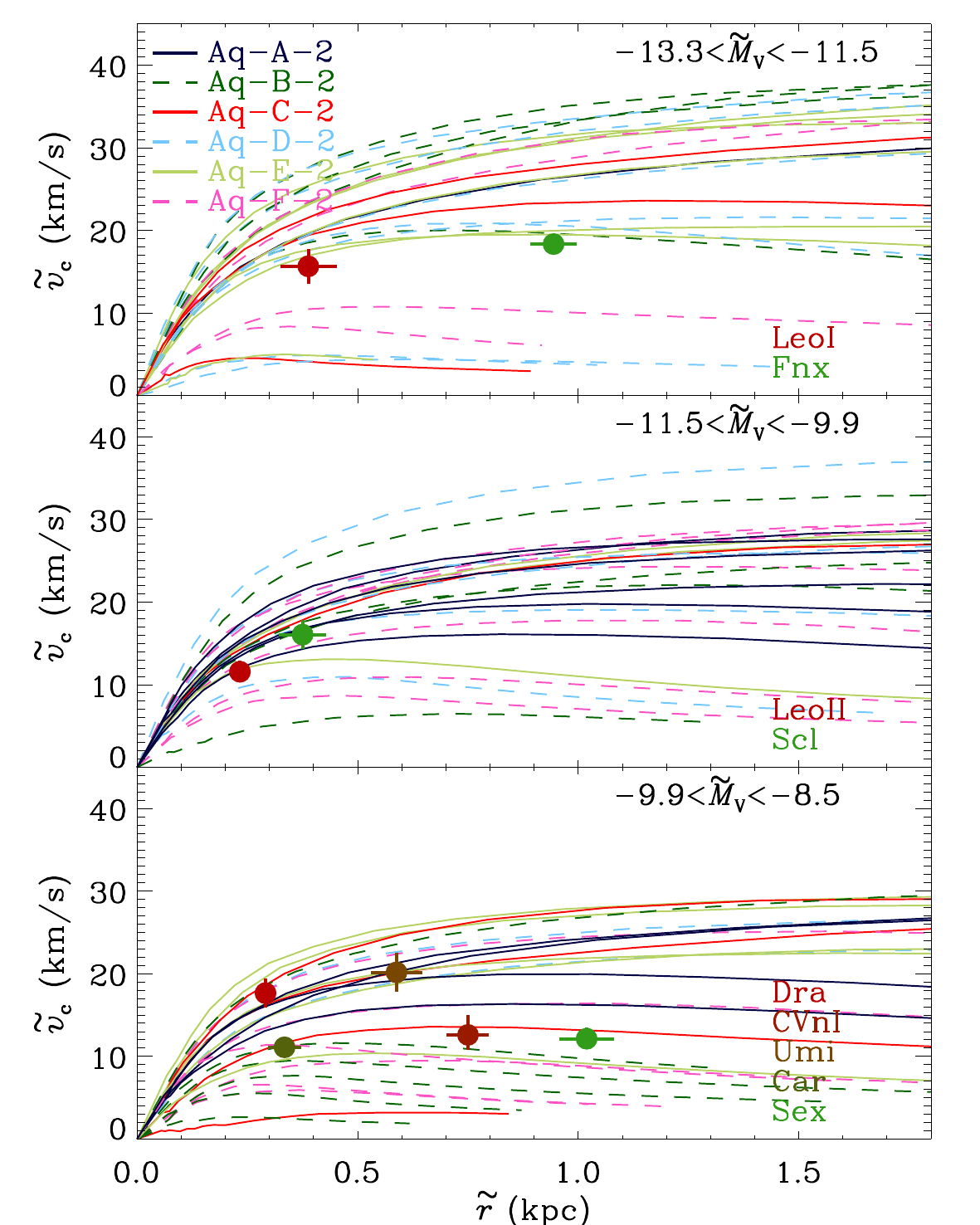}
  \caption{ Circular velocity profiles for scaled subhalos in three
    different luminosity bins, following the absolute magnitudes of
    the nine classical dSph of the Milky Way. The subhalos are colored
    according to the host halo they are associated with. This figure
    shows that the number of satellites per bin, as well as the
    velocities profiles are consistent with the measurements obtained
    for the dSph.}
\label{fig:vc-r-wolf}
\end{figure}

We now study more closely the circular velocity profiles of the
subhalos hosting satellites, since previous works have highlighted a
discrepancy between the observations and the simulations
\citep{Boylan2011,Lovell2011}.  Fig.~\ref{fig:vc-r-wolf} shows the
scaled circular velocity profiles for all the subhalos hosting
satellites with luminosities in the quoted range. As in previous
sections, we have also included the estimates for the 9 most luminous
dSphs of the Milky Way following \citet{Wolf2010}. The first
conclusion is that our semi-analytic model places (satellite) galaxies
of a given luminosity in the right mass (sub)halos, since the
amplitude of the rotation curves in all cases are consistent with
those observed. Secondly, the number of objects per luminosity bin is
in good agreement with the number observed, as established in
Fig. \ref{fig:lum-fn}. For example, in the most luminous bin (top
panel) the median number of bright satellites per halo is 3, while for
intermediate luminosities it is 4, and for the faintest considered
here, it is 4. We emphasize however that the range within a given
luminosity bin is quite broad. For example for the brightest bin, the
scaled \texttt{Aq-A-2} has just 1 satellite, while the scaled
\texttt{Aq-D-2} has 7. Such large variations are not unexpected, but
stresses that strong conclusions cannot be drawn when the number of
objects is so small as in the case of the bright end of the luminosity
function.

The galaxies shown in the fainter two bins agree quite well with the
predictions of our models. There is no systematic mismatch, with the
dSph circular velocity measurements at the half-light radius lying
close to the median velocity profile of the simulated satellites. An
apparent discrepancy is present in the most luminous bin $-13.2 \leq
\widetilde{M}_V \leq -11.9$ in the sense that there is a larger number
of subhalos with circular velocities above the measured values for Fnx
and Leo I than there is below. Nevertheless as discussed above, this
comparison is limited by statistics and affected by stochastic aspects
in the luminosity function.

We quantify this by comparing the observed value of the circular
velocity at $r_{1/2}$ for each dSph, with the probability distribution
function of $v_c$ calculated at the same radius using all the subhalos
that lie in the corresponding luminosity bin. We compute the median
$v_c$ and two percentiles of the distribution, namely a lower (15.9\%)
and an upper (84.1\%) value, which would correspond to $\pm 1\sigma$
in the case of a Gaussian. The results of this experiment are shown in
Table \ref{tab:sigma} for the different satellites. Note that here we
have translated the probability into an ``equivalent" $N \sigma$ away
from the median. The table shows that all satellites are consistent
with being drawn from the population of subhalos hosting galaxies
found in our simulations.

\begin{table}
  \begin{center}
    \newcolumntype{R}{>{\raggedleft\arraybackslash}X}%
    \begin{tabularx}{0.45\textwidth}{lRRRRR} \toprule[1.4pt]
      dSph  & $N\sigma$ away  & Observed $v_c(r_{1/2})$ & Median $v_c(r_{1/2})$ & med$(v_c)$ $-1\sigma $  & med$(v_c)$ $+ 1\sigma$  \\ \midrule[1.4pt]   
      Fnx   & $-0.72$ & 18.3 & 25.4 &  6.0 & 32.0\\
      LeoI  & $-0.73$ & 15.7 & 19.4 &  8.2 & 24.5\\\hline
      Scl   & $-0.47$ & 16.1 & 17.6 & 12.8 & 20.8\\
      LeoII & $-1.03$ & 11.6 & 14.2 & 11.9 & 16.8 \\\hline
      Sex   & $-0.22$ & 12.1 & 16.4 &  4.4 & 25.3\\
      Car   & $-0.40$ & 11.1 & 13.6 &  6.6 & 20.0\\
      Umi   & $+0.62$ & 20.2 & 16.2 &  5.8 & 22.4\\
      CVnI  & $-0.23$ & 12.6 & 16.3 &  5.1 & 23.0\\
      Dra   & $+1.03$ & 17.7 & 13.6 &  6.7 & 17.6\\\bottomrule[1.4pt]
    \end{tabularx}
  \end{center}
  \caption{\label{tab:sigma} Statistical comparison of the 9 most
    luminous classical dSphs with the simulated velocity profiles in
    Fig. \ref{fig:vc-r-wolf}.}
\end{table}

Thus far we have focused on the nine classical dSph, and have excluded
from the analysis the Sagittarius dwarf and the Magellanic Clouds. One
of the questions originally posed by \citet{Boylan2011b} is that there
may be a hidden population of very massive subhalos in the Milky Way,
since the circular velocities of the classical dwarfs are lower than
those found for the nine most massive (at infall) subhalos in any of
the Aquarius simulations. So far we have shown that our model predicts
the satellites of a given luminosity to be hosted in subhalos of the
right mass, when comparing to the classical dSph. However we also need
to explore what happens for systems brighter than Fnx, and whether we
indeed expect a missing population from our models. As expected, the
scaled Aquarius halos show a diverse number of systems with
$\widetilde{M}_V < -14$, ranging from 2 for \texttt{Aq-C-2} to 5 for
\texttt{Aq-A-2}. A simple comparison to the Milky Way satellite system
would suggest that we cannot argue that there is a population of
massive satellites that is missing.

This conclusion is also reached when considering $v_{\rm max}$ instead
of luminosity.  Although in general the most luminous objects are
hosted by the most massive subhalos, for $v_{\rm max} < 40$ km/s and
$\widetilde{M}_{V} > - 14$ there is significant scatter, and objects
as bright as Fnx are hosted in subhalos with $v_{\rm max} \sim 5 - 35$
km/s as the top panel of Fig.~\ref{fig:vc-r-wolf} shows. On the other
hand, objects brighter than Fnx are generally hosted by subhalos with
$v_{\rm max} > 40$~km/s. We find a median number of such subhalos of 2
within 280 kpc from the center for the scaled-down main Aquarius
halos, with 3 for \texttt{Aq-A-2} and none for \texttt{Aq-C-2}.

\begin{figure}
  \includegraphics[width=0.48\textwidth]{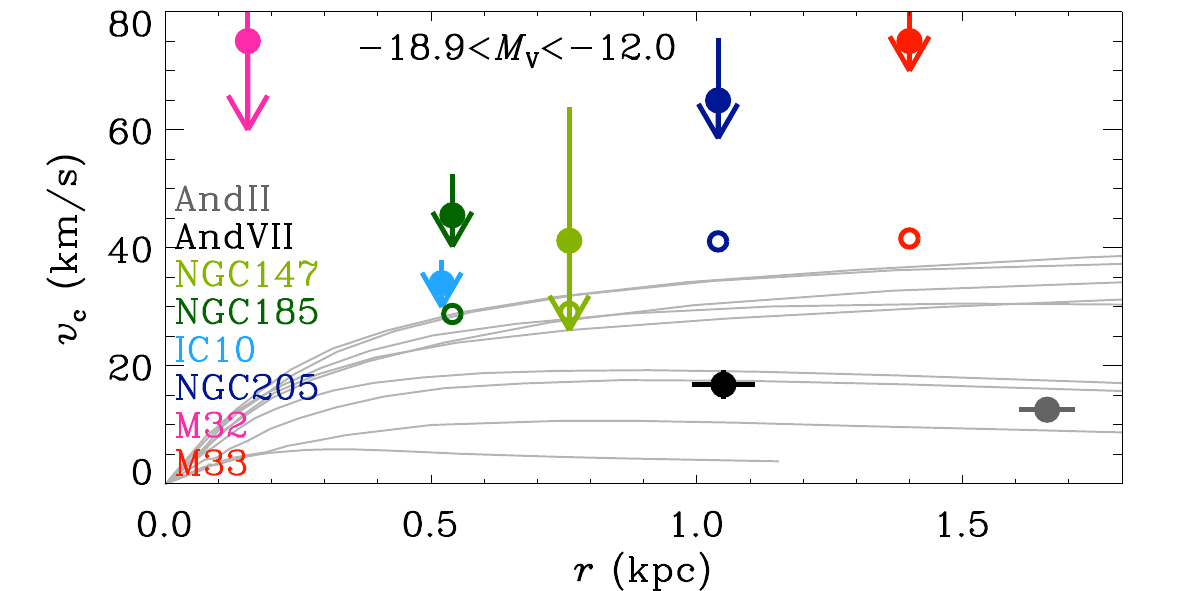}
  \caption{ Circular velocities for the subhalos present in halo
    \texttt{Aq-C-2} associated to satellites with luminosities
    $M_{V}\leq -12$. The symbols represent observations of the
    satellites of M31 in the same luminosity range. Open symbols
    represent the estimated dark matter contribution to $v_c(r)$ when
    the decomposition is available (see text for details).}
\label{fig:vc-r-geha}
\end{figure}

As suggested by multiple authors, the dark matter mass of M31 is
almost twice that of the Milky Way \citep[e.g.][]{Li2008,
  Kallivayalil2009, Guo2010}. This would imply that M31 should host
more, and also brighter, satellites than the Milky Way itself. This
indeed appears to be the case, as M31 has 8 satellites within 280 kpc
that are brighter than Leo I ($M_V = -11.9$) compared to 2 (or 5 when
Sgr and the Magellanic Clouds are included) for the Milky Way.

Aquarius halo \texttt{Aq-C-2} has $M_{\rm vir}=1.77\times 10^{12}$
M$_{\odot}$ which is $\sim 2.2$ times larger than our candidate for
the Milky Way, making this object a good match for M31. In Figure
\ref{fig:vc-r-geha} we show the velocity profiles of all the nine
satellites in \texttt{Aq-C-2} with $M_V \leq -12$. We have also
included measurements for M31's satellites with luminosities in that
range. It is important to bear in mind that these measurements have
been derived using a variety of methods that range from H\textsc{i}
rotation curves for IC10 \citep{Wilcots1998} and M33
\citep{Corbelli2003}, to 3 integral dynamical modelling for NGC147,
NGC185 and NGC2005 \cite{DeRijcke2006}. For the dSph AndII and AndVII
the method presented by \citeauthor{Wolf2010} is used to estimate
$v_{c}(r_{1/2})$ \citep{Kalirai2010}, while for M32 the mass is
derived through Jeans modelling \citep{Magorrian2001}. Many of these
bright dwarf galaxies are not as dark matter dominated within the
region populated by the stars as the dSph, and hence a direct
comparison to the circular velocity of the subhalos is not quite
correct. For example, for NGC147, NGC185 and NGC2005 the dark matter
content is estimated to be 50\%, 40\% and 40\%, respectively
\citep{DeRijcke2006}. The open symbols in Fig. \ref{fig:vc-r-geha}
correspond thus to the dark matter contribution to the circular
velocity as estimated by these authors, while the solid points
represent the total enclosed mass at the given radius. Clearly for
M32, a very compact dwarf elliptical, the shown circular velocity is
also an upper limit for the dark matter contribution.

This comparison shows that the velocity profiles for our most luminous
satellites in \texttt{Aq-C-2} are very consistent with the
observations of the dwarfs in M31 over a similar luminosity range. We
therefore must conclude there is no evidence of a missing population
of very massive of dark satellites.

\section{\amina{Discussion and Conclusions}}
\label{sec:conclusions}

We have used the state of the art cosmological $N$-body simulations of
Milky Way-like dark matter halos of the Aquarius project, supplemented
with a semi-analytic galaxy formation model, to study the dynamical
properties of the satellite population in the Local Group.

We have found that the mass profiles of the subhalos associated to
bright satellites according to our model deviate from the standard NFW
form \citep[see also][]{Stoehr2002}, and that Einasto profiles provide
much better fits. The shape parameter $\alpha$ exhibits a correlation
with the amount of mass stripped since the time of accretion,
indicating that tidal effects may be responsible for the changes in
the profiles of dark matter halos once they become satellites
\citep{Hayashi2003}.

The comparison of our models to current measurements of the mass
enclosed within the half mass radius for the classical dwarf
spheroidals suggests that they are embedded in dark matter halos of
$v_{\rm max} \sim 10-30$ km/s with $\alpha \sim 0.2-0.5$.  In
principle, this prediction for the values of the shape parameter
$\alpha$ could be tested observationally. However this requires very
extensive sampling of the kinematics of stars near the center of the
dwarf galaxies. It is also necessary to perform more sophisticated
dynamical models, that are free of assumptions regarding the velocity
anisotropy of the systems. For example Schwarzschild models of the
Sculptor dSph constrain the inner logarithmic slope of the dark matter
density profile to be $d\log \rho/d\log r > -1.5$
\citep{Breddels2011}. Better constraints could be obtained if the
sample size were increased by a factor $\sim 10$.

We have also shown that the number and internal dynamics of the
classical dSph in the Milky Way are consistent with the predictions of
the $\Lambda$CDM model, if the Milky Way's mass is $\sim
8\times10^{11}$ M$_{\odot}$. This value well within the range measured
using the dynamics of stellar tracers, but suffers from significant
uncertainties. However, it is important to note that this low value
lowers the probability of a galaxy like the Milky Way to host two
satellites as bright as the Small and Large Magellanic Clouds
\citep{busha,boylan-k}, although such systems appear to be rare in any
case, as shown by \citet{Liu} using the Sloan Digital Sky Survey.

We have also found significant scatter in the number of subhalos
expected to host bright satellites for the Aquarius halos, even when
scaled to a common mass of $8 \times 10^{11}$ M$_{\odot}$. For
example, the scaled \texttt{Aq-A-2} has five satellites brighter than
Fornax, while the scaled \texttt{Aq-C-2} has only two (making it
consistent with our Galaxy). Therefore, care should be taken to draw
strong conclusions from this region of the luminosity function since
the number of objects is small and heavily influenced by the host mass
as well as by stochastic effects associated to particular
histories. Another example that emphasizes this point is given by M31,
which plausibly is nearly a factor of two more massive than the MW,
and also has a larger number of bright satellites. Just like for the
Milky Way, our models for the satellite population of M31 are
consistent with the observational constraints on the internal dynamics
of the brighter satellites, after taking into account the differences
in host mass.

A similar conclusion was reached by \cite{Wang2012} in a paper
submitted shortly after ours, and based on the Millenium simulation
series.  These authors show that the cumulative number of subhalos
with a given peak circular velocity depends roughly linearly on host
halo mass, and that it is a highly stochastic quantity. In fact, once
normalized to the mass of the host, this function is close to
Poissonian. Based on this result, they conclude that $\sim 46\%$ of
the halos with $M= 10^{12}$ M$_{\odot}$ have no more than 3 subhalos
more massive than $v_{\rm max} = 30$ km/s, and that the percentage
increases to $\sim 61\%$ for halos with $M=M(\texttt{Aq-B-2}) =
8.2\times10^{11}$ M$_{\odot}$, in good agreement with our own
inferences. Hence we must conclude that we have found no evidence of a
missing massive satellite problem in the Local Group.

\section*{Acknowledgments}
The Aquarius simulations have been run by the VIRGO consortium, and we
are very thankful to this collaboration, and especially indebted to
Volker Springel. We are very grateful to Gabriella De Lucia and
Yang-Shyang Li in relation to the semi-analytic model of galaxy
formation, and to Simon White, Mike Boylan-Kolchin and the referee for
a critical reading of the manuscript. AH gratefully acknowledges
financial support from the European Research Council under
ERC-Starting Grant GALACTICA-240271. ES is supported by the Canadian
Institute for Advanced Research (CIfAR) Junior Academy and a Canadian
Institute for Theoretical Astrophysics (CITA) National Fellowship.

\bsp
\label{lastpage}

\bibliographystyle{mn2e}
\bibliography{refs}

\end{document}